# High Throughput Characterization of Epitaxially Grown Single-Layer MoS$_2$

Foad Ghasemi[1,2], Riccardo Frisenda[1,*], Dumitru Dumcenco[3,4], Andras Kis[3,4], David Perez de Lara[1] and Andres Castellanos-Gomez[1,5,*]

[1] Instituto Madrileño de Estudios Avanzados en Nanociencia (IMDEA-nanociencia), Campus de Cantoblanco, E-28049 Madrid, Spain.
[2] Nanoelectronic Lab, School of Electrical and Computer Engineering, University of Tehran, 14399-56191 Tehran, Iran.
[3] Electrical Engineering Institute, École Polytechnique Fédérale de Lausanne (EPFL), CH-1015 Lausanne, Switzerland.
[4] Institute of Materials Science and Engineering, École Polytechnique Fédérale de Lausanne (EPFL), CH-1015 Lausanne, Switzerland.
[5] Present address: Instituto de Ciencia de Materiales de Madrid (ICMM-CSIC), C/Sor Juana Inés de la Cruz 3, E-28049 Madrid, Spain.

*E-mail: riccardo.frisenda@imdea.org; andres.castellanos@csic.es.

**ABSTRACT:** The growth of single-layer MoS$_2$ with chemical vapor deposition is an established method that can produce large-area and high quality samples. In this article, we investigate the geometrical and optical properties of hundreds of individual single-layer MoS$_2$ crystallites grown on a highly-polished sapphire substrate. Most of the crystallites are oriented along the terraces of the sapphire substrate and have an area comprised between 10 μm$^2$ and 60 μm$^2$. Differential reflectance measurements performed on these crystallites show that the area of the MoS$_2$ crystallites has an influence on the position and broadening of the B exciton while the orientation does not influence the A and B excitons of MoS$_2$. These measurements demonstrate that differential reflectance measurements have the potential to be used to characterize the homogeneity of large area CVD grown samples.

## INTRODUCTION

Rapidly after the first works on mechanically exfoliated MoS2, part of the efforts of the scientific community working on two-dimensional (2D) materials focused on developing synthesis methods that could provide large-area single-layer MoS2 [1-6]. Chemical vapor deposition (CVD) based growth methods, which were already successfully used in the growth of graphene, are nowadays standard techniques to grow large area samples of single-layer MoS2 and other transition metal dichalcogenides [7-12]. In fact, CVD samples have shown remarkable electronic and optical properties approaching those of mechanically exfoliated material [13-16].

Although some techniques such as Raman spectroscopy and photoluminescence (PL) mapping have been used to investigate the structural and electronic properties of 2D layers based CVD growth [17-21], a thorough statistical study of the uniformity of the as-grown samples is still somewhat lacking. In this work we use micro-reflectance spectroscopy measurements to investigate the differential reflectance spectra of hundreds of CVD-grown MoS2 flakes grown on a highly-polished sapphire substrate. This technique is a very fast and non-destructive characterization tool that allows measuring a large number of spectra in different sample locations and in a small amount of time. This can provide useful statistical information about the homogeneity of the optical properties of the samples.

**MATERIALS AND METHODS**

The CVD growth of MoS$_2$ is based on the gas-phase reaction between MoO$_3$ (≥ 99.998% Alfa Aesar) and high-pure sulfur evaporated from the solid phase (≥ 99.99% purity, Sigma Aldrich), for detailed information about the growth see reference [22]. Figure 1a shows an optical microscopy image (epi-illumination mode) of a 60 x 60 μm region of the sample containing single-layer MoS$_2$ crystallites grown onto the sapphire surface. Most of the crystallites have approximately an equilateral triangular shape and they appear to be oriented following preferential orientation with respect to the sapphire lattice as previously reported [22, 23]. As can be seen in the right part of the picture, some of the crystallites can coalesce and merge. In this article we focus only on the isolated triangular MoS$_2$ crystallites for our analysis. To extract statistical information about the geometry of the MoS$_2$ crystallites, we collected 50 optical images (108 × 86 μm) of the sample surface, taken in different regions. From each image we extract the area and the orientation of the single-layer MoS$_2$ triangular crystallites.

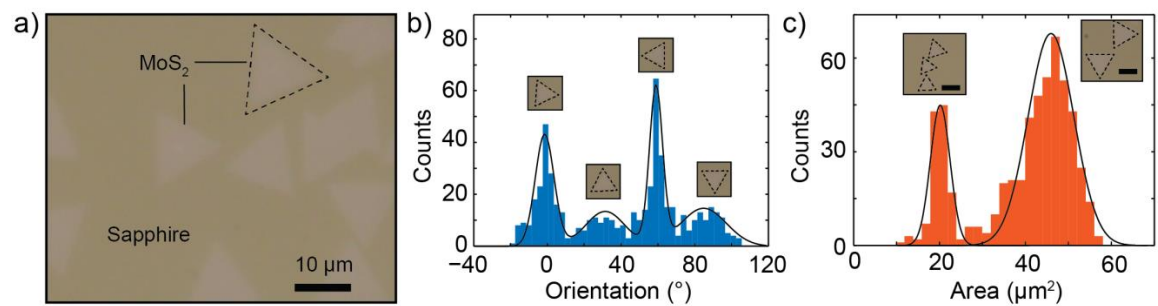

**Figure 1.** (**a**) Optical image of the grown single layer MoS$_2$ with triangle shape (the dash-line highlights single crystallites) on sapphire. (**b**) Orientation histogram of 550 MoS$_2$ crystallites. The black line is a fit to four Gaussian peaks. Inset: orientation of four crystallites corresponding respectively to an angle of 0°, 30°, 60° and 90°. (**c**) Area histogram of 550 MoS$_2$ flakes. The black line is a fit to two Gaussian peaks. Inset: optical pictures of two different regions of the sample showing respectively small-area crystallites and large-area ones. The scale bar is in both cases 10 μm.

The optical characterization of the single-layer MoS$_2$ crystallites has been carried out with differential reflectance spectroscopy. More details on the technique and about the experimental setup can be found in Ref. [24] but briefly we illuminate a 60 μm spot on the sample surface with white light and we collect the light reflected from an area of the sample of 2 μm in diameter. Subtracting from the intensity of the light reflected by the MoS$_2$ flake the intensity reflected by the bare substrate, and normalizing the result by the intensity reflected by the MoS$_2$, one can extract a differential reflectance spectrum, which in the case of a thin film is proportional to the absorption of the thin film [25, 26].

**RESULTS**

We analyzed 550 MoS$_2$ triangular flakes and we built histograms of the area and the orientation. Figure 1b shows a histogram of the triangle orientation with the orientation angle defined as in the pictures in the inset. The histogram presents two main peaks located at 0° and 60° and two secondary peaks centered at 30° and 90°. We fit the histogram to four Gaussian peaks and we find that 380 flakes (69% of the total population) are oriented around 0° or around 60° while 170 (31 %) flakes have an angle around 30° or 90°. The presence of clear peaks in the orientation histogram indicates the presence of preferential orientation populations of MoS$_2$ triangles on the surface of the sample. Since the single-crystal sapphire substrate has a two-fold symmetry (it is invariant for a rotation of 180°), the populations with orientation of 0° and 60° are physically equivalent and are the most probable ones (as well as two other orientations at 30° and 90°). The presence of these two different populations can be explained by the growth dynamic of CVD MoS$_2$ onto sapphire [23, 27] with the most probable configuration, the 0°/60°, being the one where MoS$_2$ islands and the sapphire lattice are commensurate.

Similarly to the previous analysis, we extracted the area of 550 MoS$_2$ islands and we built a histogram of the triangles area, which is shown in Figure 1c. The histogram exhibits two peaks, one centered at 20 µm$^2$ and the other at 47 µm$^2$. We fitted two Gaussian peaks to the histogram and from the peaks areas we find that 118 flakes (22% of the total population) have a small area (around 20 µm$^2$) while 431 flakes (78%) have a large area (around 47 µm$^2$). We attribute this difference in the triangle size distribution to be originated by spatial variations of the temperature and/or precursors flow during the growth process over the sapphire.

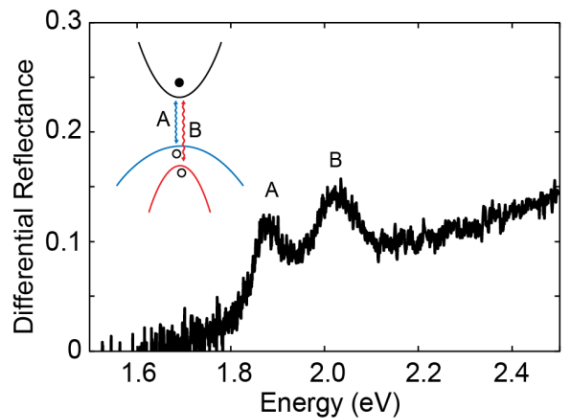

**Figure 2.** Differential reflectance spectrum of a single-layer MoS$_2$ triangle showing A and B exciton peaks. Inset: schematic drawing of the band diagram of single-layer MoS$_2$ indicating the A and B excitons at the K point.

We further characterize the homogeneity of the sample by means of micro-reflectance spectroscopy. Figure 2a shows a differential reflectance spectrum of a MoS$_2$ island. The spectrum increases for increasing energy and is characterized by two prominent peaks centered at 1.877 eV and 2.016 eV which correspond to excitonic resonances originating from direct valence to conduction band transitions at the K point [26, 28, 29], as schematically depicted in the inset of Figure 2. The presence of two features is due to the spin-orbit splitting mostly of the valance band and thus the difference in energy between these peaks gives a good estimation of the spin-orbit interaction in MoS$_2$.

For each one of the 550 MoS$_2$ crystallites previously identified, we carried out differential reflectance measurements. From each spectrum, we extracted the energy and the full-width-at-half-maximum (FWHM) values of the A and B excitons by fitting Gaussian curves on the obtained spectra. Figure 3 shows the histograms built up with those values. The energy of the A and B exciton and their broadening have variations across the CVD sample as evidenced by the finite width of all four histograms. The shape of the histograms, which is not that of a simple Gaussian curve, suggests that random fluctuations alone cannot explain the origin of the exciton variability. Possibly additional mechanisms, like the interaction with the substrate, may play a role in the spectrum-to-spectrum variability. We extract the maximum of each histogram and we find that the average A and B excitons energy is respectively 1.876 eV and 2.016 eV, while the FWHM of the A and B excitons is in average 71 meV and 124 meV respectively. Comparing these values with values that we found previously for mechanical exfoliated single-layer MoS$_2$ [24] (A 1.90 eV, B 2.04 eV) we see that an overall blueshift of 24 meV is present in both the A and B exciton position of CVD-grown MoS$_2$. The difference between the A and B mean energy is 140 meV, which is comparable with previous studies of single-layer MoS$_2$ [26, 28].

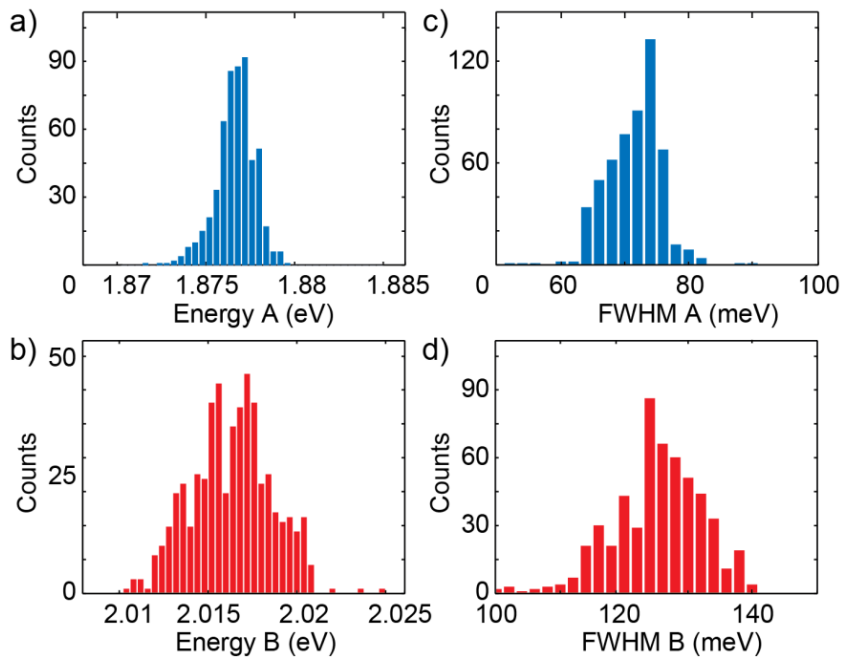

**Figure 3.** Extracted data from Differential reflectance measurement: (**a-b**) Histograms of the A and B exciton energy for 550 flakes. (**c-d**) Histograms of the full width at half maximum of the A and B excitons peaks.

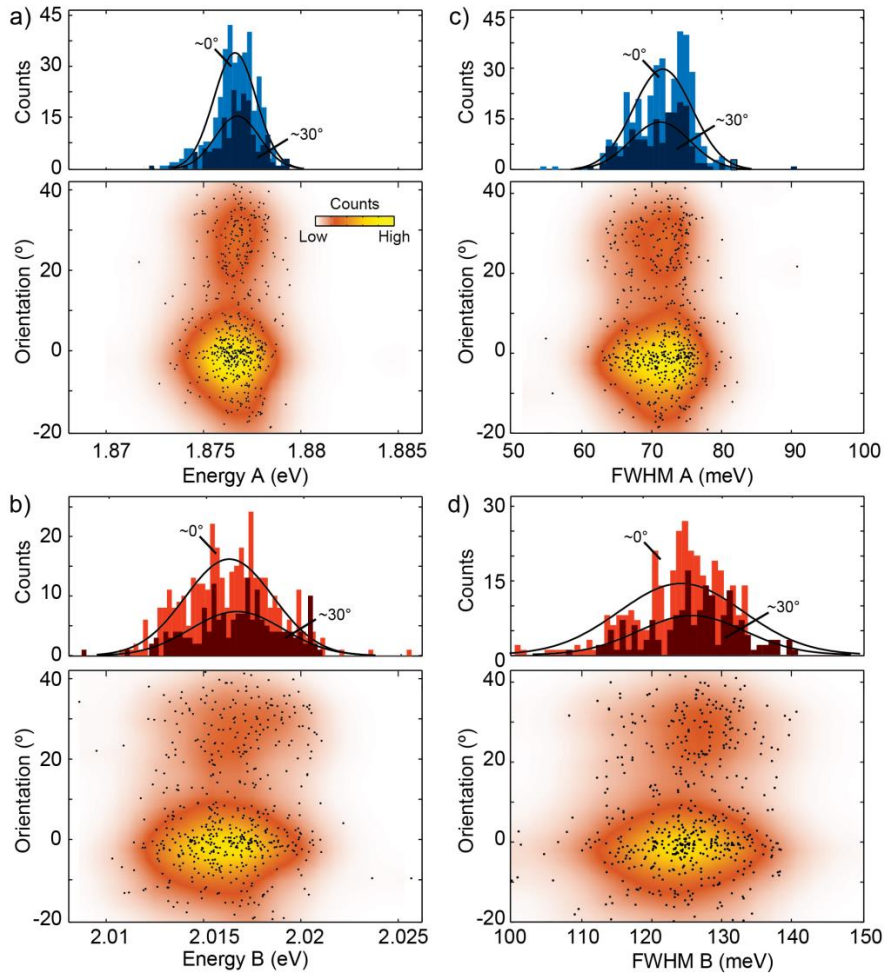

**Figure 4.** (**a**) Density plot of the triangle orientation as a function of the A exciton energy. (top) Conditional histograms of the A exciton energy for triangles with orientation 0° ± 10° (light blue) and triangles with orientation 30° ± 10° (dark blue). (**b**) Same as panel (a) for exciton B. (**c**) Density plot of the triangle orientation as a function of the A exciton peak FWHM. (top) Conditional histograms of the A exciton peak FWHM for triangles with orientation 0° ± 10° (light blue) and triangles with orientation 30° ± 10° (dark blue). (**d**) Same as panel (c) for exciton B.

In order to get a deeper insight in this spectrum-to-spectrum fluctuations observed in CVD $MoS_2$ sample, we investigate the correlations between the A and B exciton energy and FWHM and the geometrical parameters of the $MoS_2$ islands such as the area and the orientation. Figure 4a shows the correlation between the A exciton energy with the triangle orientation. The bottom plot represents a scatter plot of the A exciton energy as a function of the island orientation (black dots) on top of a color map which depends on the data points density [30]. The scatter plot shows two clouds of data points corresponding to the most probable orientation (0° or equivalently 60°) and the least probable orientation (30° or 90°). Both clouds appear to be centered at the same A energy suggesting that the orientation does not affect the excitons in CVD $MoS_2$. In the top panel we have included two one-dimensional conditional histograms of the A exciton energy, each constructed only from the islands belonging to one of the two orientation populations. The conditional histograms of the two populations appear similar and by fitting each histogram to a Gaussian peak we find that the difference in energy of the A exciton between the 0° and 30° orientation is (0.1 ± 0.1) meV, which can be considered zero within the experimental uncertainty. The uncertainty is calculated by propagating the standard errors of the peaks center. Similar conclusions hold for the energy of exciton B, presented in Figure 4b, the difference in the energy of the B exciton between triangles with 0° and 30° orientation is (0.4 ± 0.3) meV. We performed the same correlation analysis for the broadening (FWHM) of the A and B peaks. Figures 4c and 4d report similar plots for the excitons peak FWHM. Again we do not observe a correlation between the broadening of the excitons and the orientation of the MoS$_2$ islands and finding a negligible difference between the two distributions, (0.4 ± 0.5) meV in the case of exciton A and (1.5 ± 0.9) meV in the case of B. This small difference in average excitons energy for the different MoS$_2$ orientations might be due to the different MoS$_2$/sapphire interaction that could induce a different strain in the MoS$_2$ layers with the 0° and 30° orientations [11, 31].

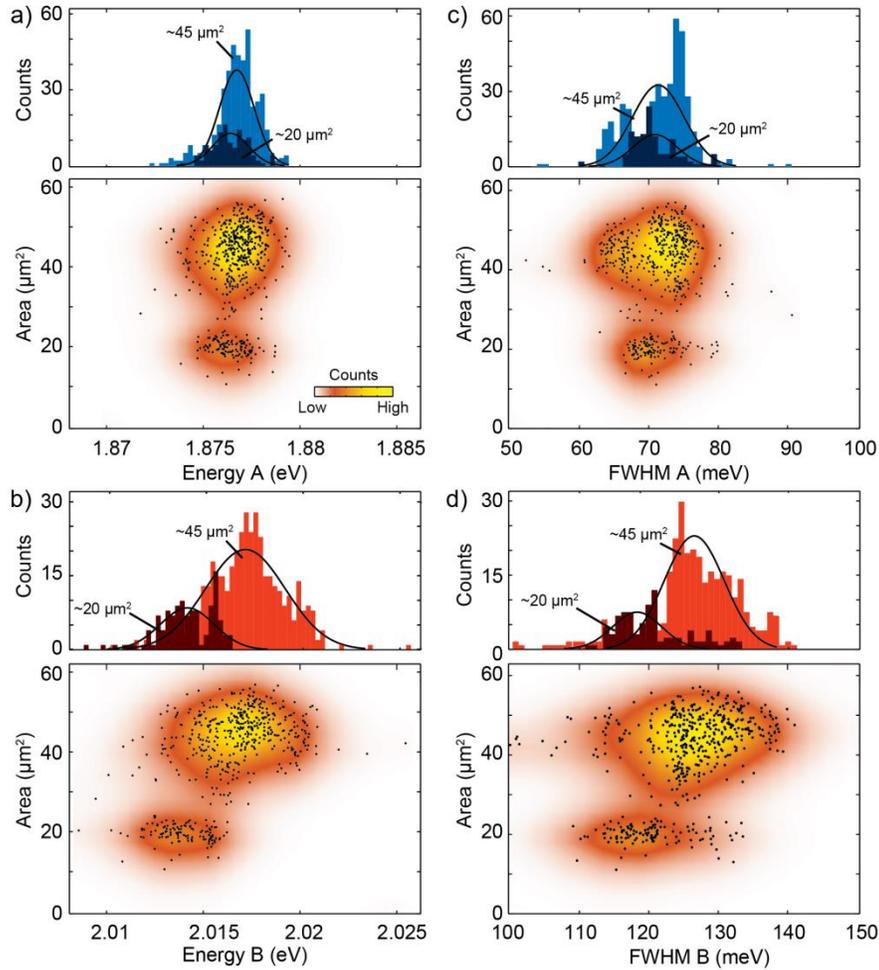

**Figure 5.** (**a**) Density plot of the triangle orientation as a function of the A exciton energy. (top) Conditional histograms of the A exciton energy for triangles with area between 15 µm² and 25 µm² (light blue) and triangles with area between 35 µm² and 55 µm² (dark blue). (**b**) Same as panel (a) for exciton B. (**c**) Density plot of the triangle orientation as a function of the A exciton peak FWHM. (top) Conditional histograms of the A exciton peak FWHM for triangles with area between 15 µm² and 25 µm² (light blue) and triangles with area between 35 µm² and 55 µm² (dark blue). (**d**) Same as panel (c) for exciton B.

Finally we perform the same correlation analysis discussed in Figure 4 to the MoS$_2$ islands area instead of the orientation. Figure 5a and 5b show the correlation between the A and B exciton energy with the triangle area. The scatter plot shows two clouds of data points corresponding to the smallest triangle area (≈20 μm$^2$) and to the largest area (≈47 μm$^2$). Interestingly, while for exciton A both clouds appear to be centered at the same energy, in the case of exciton B the points related to the smaller area are shifted toward smaller energy. In the top panel we have included one-dimensional conditional histograms of the A (B) exciton energy, which are constructed only from the islands belonging to one of the two area populations. The conditional histograms of the two populations appear similar for the A exciton energy, and by fitting each histogram to a Gaussian peak we find that the difference in energy of the A exciton, between the small and the large area populations, is (0.4 ± 0.1) meV. In the case of the B exciton instead we find that the smaller area islands have an energy (3.0 ± 0.2) meV smaller than the large area ones. We performed the same correlation analysis for the broadening (FWHM) of the A and B peaks and figures 5c and 5d report similar plots for the excitons peak FWHM. We do not observe a correlation between the broadening of the A exciton and the area of the MoS$_2$ islands finding a negligible difference of (0.5 ± 0.5) meV between the two distributions. The B exciton FWHM shows instead a marked difference of (7.5 ± 1.0) meV between large and small areas.

**DISCUSSIONS**

The results of the previous section indicate that the orientation of the MoS$_2$ crystallites in respect to the sapphire substrate does not affect the A and B excitons energy or broadening. Conversely, the area of the crystal shows a correlation with the position and broadening of the B exciton. From literature it is known that CVD-grown MoS$_2$ can show non-homogeneous strain [21, 32, 33]. One of the possible scenarios is that in our sample the CVD growth method introduces strain in the MoS$_2$ lattice, thus modifying the bands of MoS$_2$ and the excitons. The area of the MoS$_2$ crystallite can determine the amount of strain transfer with the substrate and thus influence the band structure. A full understanding of this effect is still missing, but it deserves further investigation in future works.

To conclude, we employed micro-reflectance spectroscopy to statistically study the variation in the optical properties of single-layer MoS$_2$ triangular crystallites grown by CVD on a sapphire substrate. We measured the spectra at 550 positions in the sample and we found the distributions of the A and B exciton energies and broadening. Interestingly, we found a correlation between the B exciton energy and broadening and the MoS$_2$ crystallite area. We thus demonstrate that this technique has the potential to be used to characterize the homogeneity of large area CVD grown samples.

**ACKNOWLEDGEMENTS**

A.C.G. acknowledges support from the European Commission (Graphene Flagship: contract CNECTICT-604391), the MINECO (Ramón y Cajal 2014 program RYC-2014- 01406 and program MAT2014-58399-JIN) and the Comunidad de Madrid (MAD2D-CM Program S2013/MIT-3007). R.F. acknowledges support from the Netherlands Organisation for Scientific Research (NWO, Rubicon 680-50-1515). D.P.dL. acknowledges support from the MINECO (program FIS2015-67367-C2-1-p). A.K and D.D. acknowledge funding fromSwiss SNF Sinergia Grant no. 147607.

**COMPETING INTERESTS**

The authors declare no competing financial interests.

**CONTRIBUTIONS**

F.G. collected the differential reflectance spectra and analyzed the optical images. F.G. R.F. and A.C.G. analyzed the experimental data. A.K. and D.D. performed the MoS$_2$ CVD growth. A.C.G. and D.PdL. designed the experiment. All the authors contributed to discussions and writing of the manuscript.